%% file: ms_astro-ph.tex
\documentclass[12pt,preprint]{aastex}

\begin{document}

\slugcomment{{\it Accepted for publication in ApJS on 12 February 2007.}}

\title{CRATES: An All-Sky Survey of Flat-Spectrum Radio Sources}

\author{Stephen E. Healey\altaffilmark{1,8}, Roger W. Romani\altaffilmark{1,8},
Gregory B. Taylor\altaffilmark{2},\\ Elaine M. Sadler\altaffilmark{3},
Roberto Ricci\altaffilmark{4}, Tara Murphy\altaffilmark{3,5}, James S. Ulvestad\altaffilmark{6},
Joshua N. Winn\altaffilmark{7}}

\altaffiltext{1}{Department of Physics/KIPAC, Stanford University, Stanford, CA 94305, USA}
\altaffiltext{2}{Department of Physics and Astronomy, University of New Mexico, Albuquerque, NM 87131, USA}
\altaffiltext{3}{School of Physics, University of Sydney, NSW 2006, Australia}
\altaffiltext{4}{Australia Telescope National Facility, CSIRO, Epping, NSW 1710, Australia}
\altaffiltext{5}{School of Information Technologies, University of Sydney, NSW 2006, Australia}
\altaffiltext{6}{National Radio Astronomy Observatory, Socorro, NM 87801, USA}
\altaffiltext{7}{Department of Physics, Massachusetts Institute of Technology, Cambridge, MA 02139, USA}
\altaffiltext{8}{Email: {\tt sehealey@astro.stanford.edu, rwr@astro.stanford.edu}}

\begin{abstract}
\ We have assembled an 8.4~GHz survey of bright, flat-spectrum ($\alpha > -0.5$) radio sources
with nearly uniform extragalactic ($|b|>10^\circ$) coverage for sources brighter
than $S_{4.8 \mathrm{\:GHz}} = 65$~mJy.  The catalog is assembled from existing observations
(especially CLASS and the Wright et al.\ PMN-CA survey), augmented by reprocessing of archival
VLA and ATCA data and by new observations to fill in coverage gaps.  We refer to
this program as {\bf CRATES}, the {\bf C}ombined {\bf R}adio {\bf A}ll-sky {\bf T}argeted
{\bf E}ight~GHz {\bf S}urvey. The resulting catalog provides precise 
positions, sub-arcsecond structures, and spectral indices for some 11,000 sources. 
We describe the morphology and spectral index distribution of the sample and
comment on the survey's power to select several classes of interesting sources,
especially high energy blazars. Comparison of CRATES with other high-frequency
surveys also provides unique opportunities for identification of high-power radio
sources.
\end{abstract}


\section{Introduction}

	As extrema of the AGN population, blazars are of particular interest
for a number of topics in accretion and jet physics. These sources are characterized
by flat radio spectra; high variability, especially in the optical; significant
polarization; and bimodal synchrotron/Compton SEDs.  They are believed to be 
high-power radio AGN with a strong jet component viewed ``pole-on." For this model,
Doppler boosting ensures that non-thermal jet emission will be strong. In
cases where the non-thermal emission dominates the thermal flux from the accretion
and surrounding broad-line region,
the objects are known as HBL or BL Lacs. We are particularly interested
in blazars since, at $\gamma$-ray energies, high-power blazars appear
to dominate the observed EGRET sources \citep{hart}.  The GeV Compton
peak flux, in fact, likely dominates the cosmic background radiation at these
energies.  Similarly, the synchrotron IR-mm peak can dominate the
point source contribution to the microwave sky \citep{giommi}. Thus, 
large blazar surveys, probing the blazar population and its evolution, can
be helpful for both $\gamma$-ray and microwave source identifications and
for understanding cosmic backgrounds in these energy bands. 

        Flat-spectrum radio surveys are the prime source of blazar
discoveries. Moreover, studies show that bright, flat-spectrum sources
strongly correlate with sources in the $>$100~MeV sky \citep{mat,hart,srm}.
In particular, our work has shown that high-energy associations are
especially powerful when interferometric measurements of core
flux density and spectral index are available.  Further, since the ``trough"
between the blazar radio and $\gamma$-ray components lies in
the optical to X-ray range, counterparts at these wavelengths are often
faint.  Indeed, many are known to have $R>23$, well below the sensitivity of
the Second Digitized Sky Survey (DSS2).  Positive identification is greatly
helped by precise sub-arcsecond
positions and structures, which require interferometric measurements at
cm wavelengths.

	Our survey is designed as an extension of the largest high-frequency
interferometric survey currently available, the Cosmic Lens
All-Sky Survey (CLASS) \citep{class}. We have replicated as closely as
possible its selection criteria and extended the survey to the full sky
at high Galactic latitudes through a combination of published data,
reanalysis of archival data, and new observations. The product represents
the largest sample of bright, compact, flat-spectrum sources available to date.

\section{Sample Selection}

\subsection{Finding Sources}

\ The basis for the CLASS sample selection was the GB6 catalog \citep{gb6}
of sources in the declination range $0^\circ < \delta < +75^\circ$ measured at 
4.85~GHz with the erstwhile 91~m telescope at Green Bank.  In order to be observed 
as part of CLASS, a source had to lie outside the Galactic plane ($|b| > 10^\circ$) 
and have a GB6 flux density of at least 30~mJy, at least one 1.4~GHz NVSS \citep{nvss}
source within $70\arcsec$ of the GB6 position, and a spectral index 
$\alpha > -0.5$ (where $S_\nu \propto \nu^\alpha$) computed between GB6 and NVSS.  
In the common case of multiple NVSS sources within $70\arcsec$ of a single GB6 position, 
the flux densities of the NVSS sources were added together, and this sum and the 
GB6 flux density were used to obtain the spectral index.  All sources that 
satisfied these criteria were then observed interferometrically at 8.4~GHz with the 
VLA in the ``A" configuration.

\ In order to produce a CLASS-like all-sky survey, we attempt to reproduce the CLASS 
selection criteria as closely as possible with the exception of the 4.8~GHz flux 
density threshold, which we increase to 65~mJy.  Consequently, in the CLASS region 
itself ($0^\circ < \delta < +75^\circ$, $|b| > 10^\circ$), CRATES is 
almost entirely a subset of CLASS.

\ Outside of the CLASS region, there are no GB6 observations, so it is necessary to 
introduce other catalogs as surrogates in order to select the sample.  Single-dish 
observations of the southern sky at 4.85~GHz are available from the PMN survey 
catalog \citep{pmn}, which covers the region $-87^\circ < \delta < +10^\circ$.  Thus, 
below the equator, PMN serves as the parent survey for CRATES in the same way that 
GB6 serves for CLASS.  To compensate for the different size of the dish used to conduct 
the PMN survey (the 64~m telescope at Parkes), we increase the matching radius for 
finding NVSS counterparts to $110\arcsec$. Further, NVSS observations are only available for 
$\delta > -40^\circ$, so below this declination, we substitute the 2006 June~1 
version of the 843~MHz SUMSS catalog \citep{sumss1,sumss2} as our low-frequency 
survey for determining spectral indices.  In this southernmost 
region of the sky, then, the sample selection is 
determined by PMN and SUMSS. Component matching in the low-frequency survey and the
spectral index and flux density cuts are made to match the selection in the north.

\ GB6 is also unavailable in the north polar cap ($\delta > +75^\circ$), 
so sources there must be drawn from other high-frequency surveys.  In this region, our
prime source is the S5 catalog \citep{s5} of sources in the range $+70^\circ < \delta 
< +90^\circ$ observed at 4.85~GHz with the Effelsberg 100~m telescope.  This catalog 
is only complete to 250~mJy, so CRATES in the northern cap is considerably shallower 
than over the rest of the sky.  NVSS does cover this region, though, so the 
CLASS-style matching procedure and spectral index cut can be applied to 
determine this part of the CRATES sample.  A summary of this patchwork of surveys 
is shown in Table 1; in the rest of this paper, we will refer to the four sky 
regions by the names shown in the table.

\begin{deluxetable}{cccccccc}

\tabletypesize{\scriptsize}
\tablecaption{General characteristics of the CRATES sample.}
\tablewidth{0pt}
\tablehead{

  \colhead{Name of}&
  \colhead{Declination}&
  \colhead{4.85 GHz}&
  \colhead{4.85 GHz flux}&
  \colhead{Low frequency}&
  \colhead{Number of}&
  \colhead{Source density}\\

  \colhead{sky region}&
  \colhead{range}&
  \colhead{survey}&
  \colhead{density depth}&
  \colhead{survey}&
  \colhead{sources}&
  \colhead{(number / $\sq^\circ$)}}

\startdata
Far North&$+75^\circ < \delta < +90^\circ$&S5&250 mJy&NVSS (1.4 GHz)&79&0.111\\
North (CLASS region)&\phs\phn0$^\circ < \delta < +75^\circ$&GB6&65 mJy&NVSS (1.4 GHz)&4886&0.298\\
Equatorial South&$-40^\circ < \delta <$\phm{ $+0$}$0^\circ$&PMN&65 mJy&NVSS (1.4 GHz)&4106&0.362\\
Far South&$-87^\circ < \delta < -40^\circ$&PMN&65 mJy&SUMSS (0.84 GHz)&2060&0.368\\

\enddata
\end{deluxetable}

\subsection{Uniformity of the Sample}

\ In order to ensure that the CRATES sample selection is as uniform as possible, we
must cross-calibrate the several surveys' flux density measurements. Of course, the
sources present in multiple surveys were not observed simultaneously, and
blazars are significantly variable, so our cross-calibration can only
be statistical. However, the mean flux density ratios give an estimate of any
flux density scale offsets while the dispersions can be corrected for the measurement
uncertainties to give an estimate of typical source variability.  Additional
variability estimates are also sometimes available from repeated observations 
within a given survey. For the 4.85~GHz catalogs, we identified common sources
in the overlap zones.  For example, GB6 and PMN overlap at
$0^\circ < \delta < +10^\circ$, with $\sim$3000 common sources. The flux density
ratio distribution shows that these surveys' flux density scales are consistent
within the expected statistical and variability fluctuations.  Similarly, over 200
sources are common to the S5 and GB6 surveys; again, there is no significant offset.
Thus, all 4.85~GHz selection is on a consistent flux density scale.

\ The situation is more complex for the low-frequency surveys used for
spectral index selection. NVSS and SUMSS have a substantial overlap region
($-40^\circ < \delta < -30^\circ$) but are at quite different
frequencies. Since we wish to select SUMSS sources that satisfy the
$\alpha_{4.85/1.4} > -0.5$ cut, we must use the overlap sources to set
an equivalent flux density/spectral index cut at 0.84~GHz. To set an
appropriate threshold, we compare the PMN/NVSS and PMN/SUMSS spectral
indices for $>$$10^3$ overlap sources as shown in Figure 1 along with a linear least
squares fit.  Clearly, the indices are not strictly proportional.  The observed
scatter is much larger than that due to flux density measurement errors and the
observed high-frequency variability, so we infer a non-negligible curvature in the 
radio spectra that varies from object to object.

We accordingly estimate the equivalent 4.85~GHz/1.4~GHz spectral index as \linebreak
$\alpha_{4.85/1.4}^\mathrm{equiv}~=~A\alpha_\mathrm{4.85/0.84}~+~B$, 
where $\alpha_{4.85/0.84}$ is the
observed PMN/SUMSS spectral index\linebreak and $A = 1.143$ and $B = -0.105$ are the parameters of
the fit.  This transformation was performed for all PMN/SUMSS sources in the Far South.

\begin{figure}[ht]
\begin{center}
\includegraphics[width=0.5\textwidth]{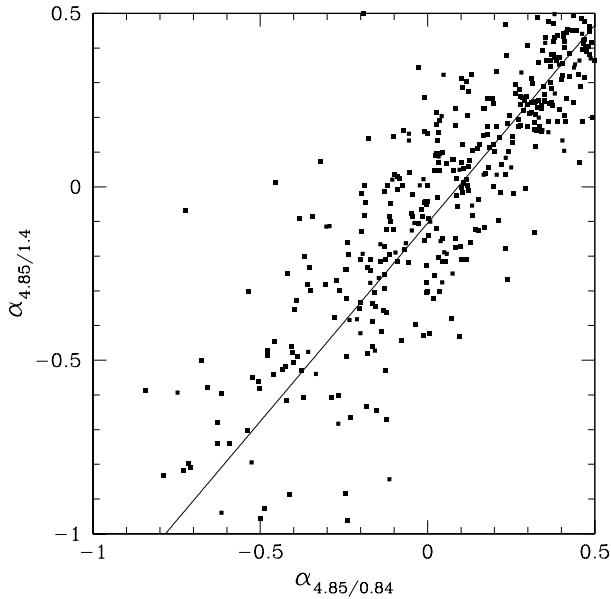}
\caption{Comparison of $\alpha_{4.85/0.84}$ and  $\alpha_{4.85/1.4}$ spectral 
indices.  The fit is $\alpha_{4.85/1.4}^\mathrm{equiv} = A\alpha_\mathrm{4.85/0.84} + B$,
with $A = 1.143$ and $B = -0.105$.}
\label{lowfreq}
\end{center}
\end{figure}

Applying the 4.85~GHz flux density cut, the $\alpha > -0.5$ spectral index 
cut (adjusted in the SUMSS region), and the Galactic plane cut, we obtained the 
final CRATES sample of 11,131
objects requiring X-band (8.4~GHz) measurements.  An Aitoff equal-area projection of the 
sample is shown in Figure 2. Several features deserve comment. The source areal
density is smaller in the Far North because of the shallower S5 survey. In the Equatorial
South and Far South, the areal densities are larger than for the North (Table 1).
This is apparently a consequence of the larger PMN 4.8~GHz beam, which will allow
more extended and composite sources to pass the 65~mJy threshold. In addition, the
presence of the Galactic bulge and the Magellanic clouds in the southern sky will introduce
some thermal flat-spectrum, but extended, sources.  Of course, composite and thermal
sources will not be truly compact and will be unmasked by the 8.4~GHz interferometric
observations. Indeed, there are more ``no-shows" in the southern sky, and the final areal
density of sources confirmed by our survey to be bright ($>$0.1~Jy 
at 8.4~GHz) is slightly {\it higher} in the North than in the two southern regions.

In addition,
there are several holes in the 4.8~GHz samples in the Equatorial South and Far South, especially
just south of $\delta = 0^\circ$. We have used lower-frequency surveys to select
nominally flat-spectrum radio candidates in these regions (see ``Additional Observations" below)
and have obtained X-band measurements, as resources permit, of the best sources.
These sources may be promoted to full CRATES status by single dish 4.85~GHz
observations. There is one obvious patch with a high density of flat-spectrum
sources: this is the Large Magellanic Cloud (LMC), and much of the excess areal density
is likely due to thermal sources in the LMC itself, but, of course, background blazars
are present.  The 8.4~GHz follow-up in this zone may be used to estimate the blazar
fraction and to explore issues in extending CRATES to the Galactic plane region.

\begin{figure}[ht]
\begin{center}
\includegraphics[width=0.8\textwidth,angle=0,keepaspectratio=true,trim=1.4in 0.8in 1in 3.5in]{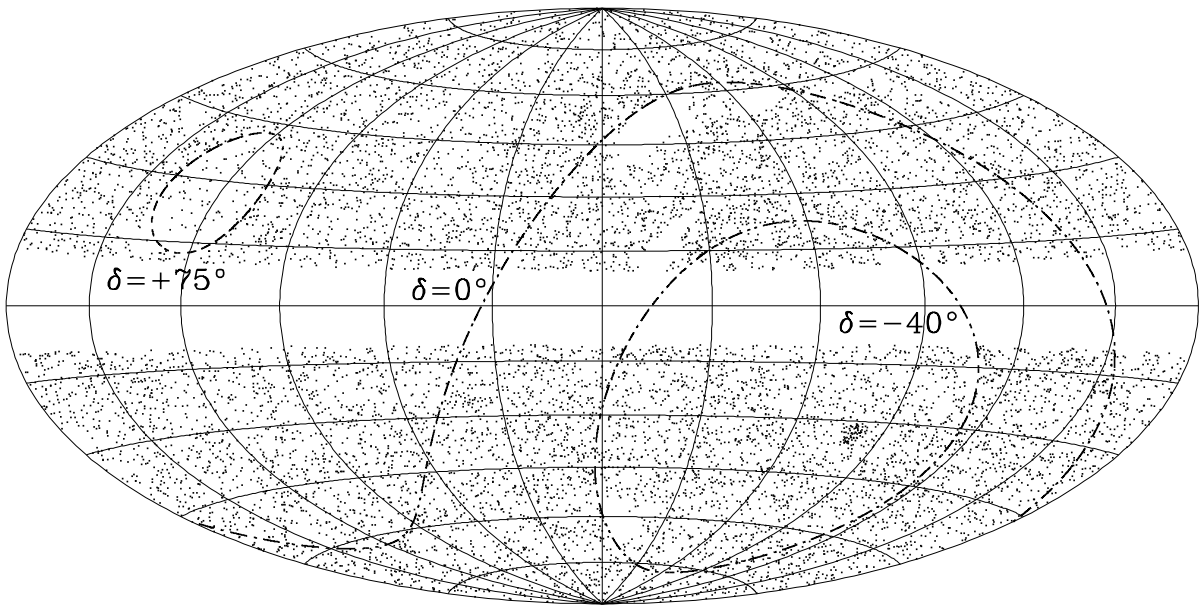}
\caption{Aitoff equal-area projection of the CRATES sample in Galactic coordinates $(l,\,b)$.  The central 
meridian is $l = 0^\circ$. The concentration of points near $\delta = -70^\circ$ 
is due to sources in the Large Magellanic Cloud.}
\label{aitoff}
\end{center}
\end{figure}

\section{X-Band Observations}

\ Fortunately,
most of the required X-band observations are available in existing observatory archives.
In particular, reduced measurements from CLASS at $0^\circ < \delta < 75^\circ$ 
cover $>$99.5\% of the CRATES-selected sources in this region. There are a few CLASS sources
observed outside this declination range.  The Equatorial South was surveyed for gravitational
lenses in a manner very similar to CLASS using PMN/NVSS selection criteria very similar to
CRATES \citep{wlens00}.  Thus, a substantial
fraction of the required X-band observations was present in the VLA archives
from a number of observing campaigns; we refer to these as CRATES-Va (VLA archive) below.

These data were remapped (and in some cases re-calibrated; see below), and measurements
of component positions and flux densities were extracted for this survey. In the Far
South, a significant source of archival observations was the
(largely unpublished) PMN-CA survey \citep{wright}; these were again remapped
and remeasured for the present survey.  A number of sources have also been measured in 
the AT 20~GHz (AT20G) Survey \citep{at20g,at20g2}, which is still in progress.  For 
the CRATES sample, we used a pre-release version of the AT20G catalog provided by 
the AT20G team and covering the declination region $-30^\circ$ to $-87^\circ$.

Some of these AT20G sources are also members of other sub-samples 
and helped to cross-calibrate the southern surveys.
To complete CRATES, we have mounted several
VLA campaigns, focusing on sources in the Far North and Equatorial South
(program numbers AR0517 = CRATES-V1, AR0555 = CRATES-V2, AR0587 = CRATES-V3).  Finally,
we mounted an ATCA campaign (program C1468 = CRATES-CA) to observe the remaining
sources south of $\delta = -40^\circ$. A handful of sources either ended up
with only one ATCA scan or had SUMSS measurements published after our final 
observing campaign and thus missed being scheduled.  The combination of the X-band data 
sets that we have collected provides $>$99.9\% coverage of the CRATES targets.

\subsection{New VLA and ATCA Campaigns}

	The three new VLA observational campaigns were each conducted with the VLA
in the ``A" array, using two 50~MHz bands at 8.44~GHz, but the observing schedule differed
somewhat between runs. The first campaign, AR0517 (2003 July~23-25), targeted
likely counterparts of southern 3EG $\gamma$-ray sources and is described in \citet{rsmu}.
The second campaign, AR0555, took place on 2004 October~2-3 and 2004 October~9. Here, some
1050 scans (duration 60-120~s) of over 900 sources were made. Standard AIPS
calibration was performed followed by DIFMAP imaging and Gaussian component fitting.
The final clean-up campaign, AR0587 (2006 March~22, 2006 April~1, and 2006 April~3), targeted
some 282 survey sources plus frequent calibrator visits, obtaining $\ge$50~s
on each. Again, standard mapping, calibration, and Gaussian component fitting were
performed.

	The ATCA campaign, program C1468, observed for 45~h on 
2005 December~22-23 using the ``6A" array and the dual frequency 6~cm/3~cm 
(4.8~GHz/8.6~GHz) system. Some 750 program targets plus calibrators 
were observed. We attempted to schedule three 60~s ``cuts" for each 
target at HA = $-4$~h, 0~h, $+4$~h for the best possible $uv$ coverage. While most sources
did get three cuts, a number of the more northern ($\delta > -40^\circ$) targets only
had two cuts, and a few received only a single observation. We were able to
re-image all but one of the one-cut sources and several of the two-cut
sources during the 2006 VLA clean-up campaign.  The ATCA data were calibrated and mapped with
the MIRIAD package, guided by the SUMSS survey sources positions. Gaussian components 
were measured with IMSAD. We report here only on the 3~cm measurements.

\subsection{Astrometry}

\ To assess the quality of the astrometry, we identified sources 
observed at multiple epochs and compared positions of the Gaussian fit
components. To ensure that the same components are being compared at multiple
epochs, we restrict the comparison to the brightest component in each field,
provided that the component has $S_{8.4\:\mathrm{GHz}}>30$~mJy and is $\ge$$1.5\times$ brighter
than the second brightest component in the field. This avoids cases in which
map noise or variability switches the ranking of comparable components.

\ For the VLA observations, we found an RMS positional error of 
$\sigma_\mathrm{VLA} \approx 0\,\farcs06$, consistent with calibrator
positional uncertainties and the expected astrometric accuracy
under good conditions.  However, there was a significant tail of
large ($>$$0\,\farcs5$) offsets.  These were dominated by observations in a fraction of the
CRATES-Va data, program AP0282, which had relatively few calibrators and poor phase
stability.  We significantly improved the situation by recalibrating the observing
runs in AP0282 using refined positions for the calibrators and by supplementing
the calibration with strong sources for which good positions were available from the
VLA calibrator manual.  This had the effect of increasing the density of calibrator
scans in the AP0282 runs, thereby reducing phase errors on the remaining target sources.
Nevertheless, the
astrometric stability of this run was still noticeably poorer than the rest of
the data (Figure 3, left panel).  In particular, of the 304 sources common to AP0282 and the
remainder of the CRATES-Va observations, 28 show offsets of $\ge$$0\,\farcs3$, 18 are
$\ge$$1\arcsec$, and 16 are $\ge$$5\arcsec$. Clearly, while the majority
of the astrometry is quite adequate for, e.g., optical source identification,
$\sim$5\% show major (i.e., $>5\arcsec$) astrometric errors.  As a result, we discard AP0282
measurements when other observations are available. There are, however,
139 CRATES sources observed only in this program.  These are flagged in the source
catalog. Of these, $\sim$10 are expected to have erroneous positions.

	After rectification of the AP0282 data, we find that a small but significant
number of single-component sources with multi-epoch data still had
significant $\ge$5$\sigma$ (but mostly $\la$$1\arcsec$) discrepancies between 
the different measurements of the source position. In each case, one epoch's position 
had much better agreement with the NVSS source position, and so we adopted these 
component locations. The discrepant positions indicated that almost all of the
errors stemmed from the CRATES-V2 data set, which could have up to 10\% of the
images with small position errors. While we cannot effect the same test with single
pointings, as it happens, the CRATES-V2 data were the sole observation for only 
eight targets, and so we do not expect more than $\sim$1 erroneous position
from this error source.

We performed a similar analysis comparing the PMN-CA sources with the 
AT20G measurements in the South (Figure 3, right panel).  Here, the decreased
array resolution and calibrator astrometric precision result in a $\sim$$10\times$
higher RMS disagreement of $0\,\farcs56$.  Again, a tail of larger discrepancies is
present (Table 2). These results are consistent with the quoted positional accuracy of
the two surveys. The AT20G survey used a more compact ATCA configuration than PMN-CA,
with a maximum baseline of 1.5~km, and therefore has lower angular resolution and
slightly lower positional accuracy.  Wright et al.\ (1997) quote a typical
position error of $0\,\farcs6$ in each coordinate for the PMN-CA data while
the AT20G data have a median positional error of $1\,\farcs3$ in right ascension and
$0\,\farcs6$ in declination (Sadler et al.\ 2006).

As in the North, we can use the low-frequency data to referee the position 
measurements and we find that 19\% of AT20G positions differ by more than
$2\,\farcs0$ from the PMN-CA and low-frequency estimates.  Of the 63 sources
in our sample that have position measurements from AT20G alone, we therefore
expect that about 12 may have residual position errors at the 2\,$\arcsec$ level.

\begin{figure}[ht]
\centering
\plottwo{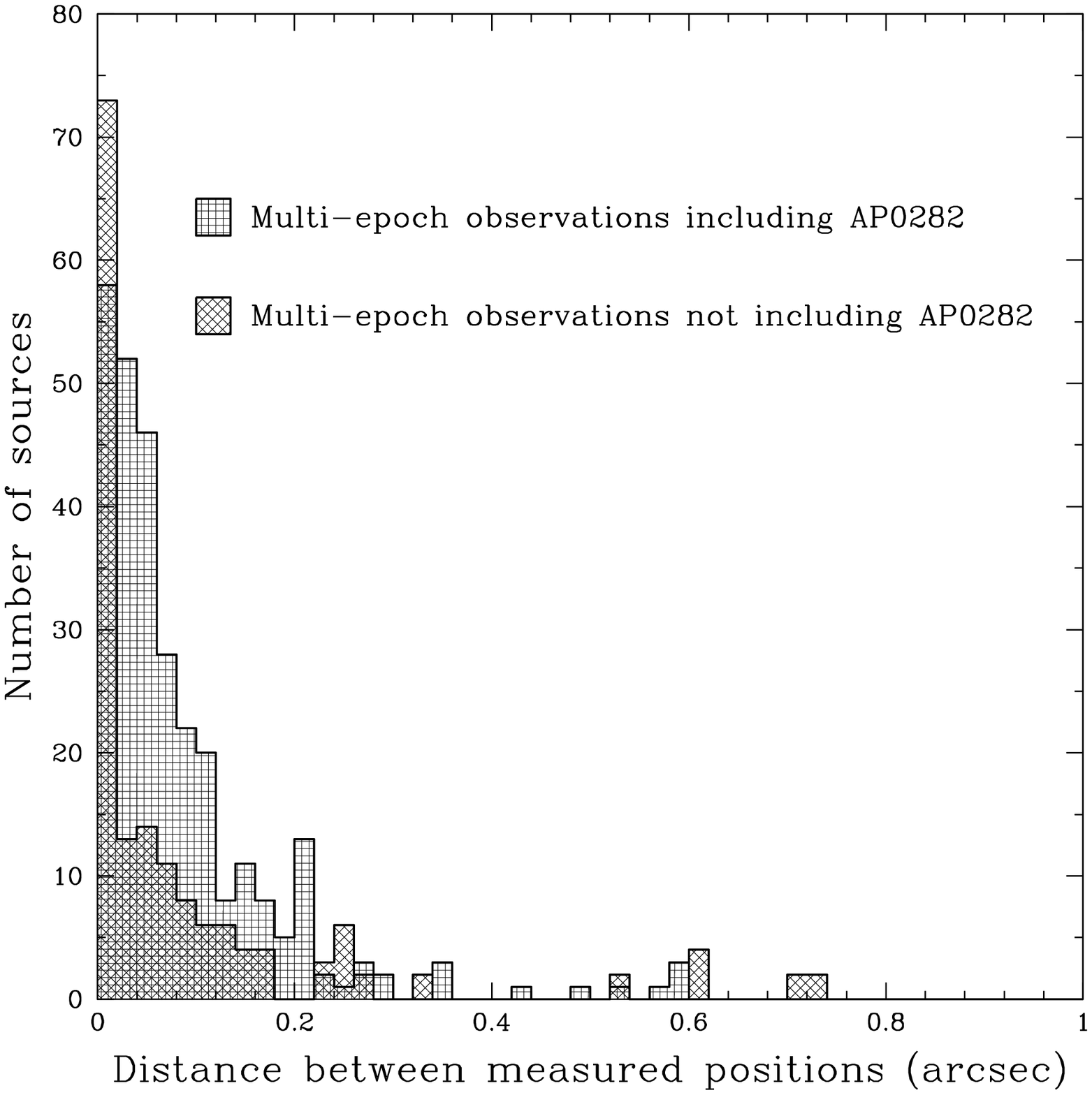}{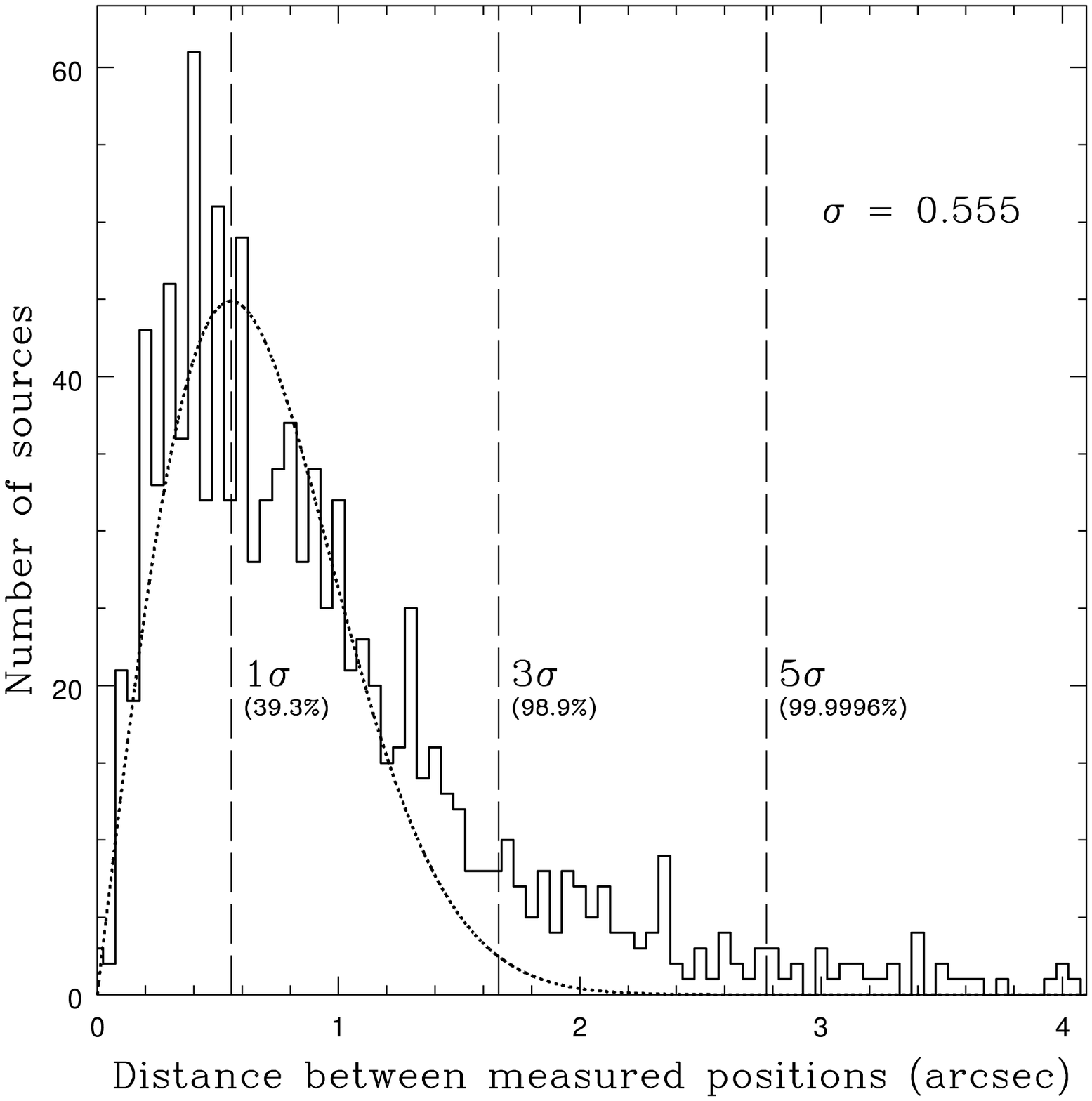}
\label{poshists}
\caption[poshists]{Astrometric cross-comparisons for the VLA and for ATCA. \\
{\it Left:} Position offsets for CRATES-Va sources observed at two epochs,
isolating those from data set AP0282, which had less stable astrometry.
See text for description of source counts extending beyond $1\arcsec$. \\
{\it Right:} Position offsets $\theta$ between PMN-CA and AT20G.  The fit is
$F(\theta) \propto \theta \, \exp{\left[-\frac{1}{2}\!\left(\frac{\theta}{\sigma}\right)^2\right]}$.}
\end{figure}

\begin{deluxetable}{cccccc}
\tabletypesize{\small}
\setlength{\tabcolsep}{0.15in}
\tablewidth{0pt}
\tablecaption{Comparative astrometry of representative surveys.}
\tablehead{
  \colhead{Comparison}&
  \colhead{$N$\tablenotemark{a}}&
  \colhead{$\sigma$\tablenotemark{b}}&
  \colhead{$\theta > 3\sigma$}&
  \colhead{$\theta > 5\sigma$}&
  \colhead{$\theta > 5\arcsec$}}
\startdata
CLASS vs. CRATES-Va &208&$0\,\farcs06$&26 (13\%)&15 (7\%)&0 (0\%)\\
AP0282 vs. rest of CRATES-Va&304&$0\,\farcs06$&54 (17\%)&28 (9\%)&16 (5\%)\\
PMN-CA vs. AT20G&1039&$0\,\farcs56$&164 (15\%)&59 (5.7\%)&24 (2\%)\\
\enddata
\tablenotetext{a}{Number of sources common to the surveys.}
\tablenotetext{b}{RMS of the fit to the distribution of offsets.  See Figure 3
for the functional form.}
\end{deluxetable}

\subsection{Radiometry}

\ We have made a similar cross-comparison of the various survey measurements to
form a uniform 8.4~GHz flux density scale, referenced to the CLASS flux densities.
Two issues complicate the comparison. First, source variability broadens the flux
density ratio histograms, so again, the flux density scale comparisons are
statistical. Second, all ATCA measurements were made at 8.6~GHz. Thus, before
cross-comparison, we project the ATCA flux densities to 8.4~GHz using the measured
PMN/SUMSS or PMN/NVSS spectral index.  This is generally a small correction; even
for a highly inverted spectrum ($\alpha = 1$), the flux density is adjusted by $<$3\%.

\ The left panel of Figure 4 shows an example cross-comparison (between CLASS and CRATES-Va)
with a log-normal fit to the distribution.  The mean $\mu$ and RMS $\sigma$
of the fit are also shown. We are interested in the uncertainty in $\mu$:
$\sigma_\mu = \sigma/\sqrt{N}$, where $N$ is the number of common sources.  For 
this particular case, $\mu = -0.0161 = -2.9\sigma_\mu$; multiplying CRATES-Va 
flux densities by $10^{-\mu} = 1.038$ brings them into agreement with CLASS.  
These corrected flux density measurements go into the final CRATES catalog.  
As it happens, CRATES-Va provides the only significant overlap with 
CLASS, so all other surveys must be referenced through this set. For example, our 
new VLA data overlap appreciably only with CRATES-Va, so they must first be
referenced to CRATES-Va in order to be brought into line with CLASS.  The results of
the head-to-head flux density comparisons are shown in Table 3.

\ Thus, for each survey, we can infer a final correction factor (CF) to
bring the flux density into agreement with the VLA CLASS measurements.  Often this
is indirect.  In the chain of corrections (Table 4), when $|\mu| < 2\sigma_\mu$, 
we conclude that there is no significant discrepancy, and the contribution 
of that particular link is set to unity in computing the CF.
In the case of AT20G, there are two possible chains; we use the one with lower 
total uncertainty in determining the CF.  In sum, the VLA flux density scales all
agree with each other and with the PMN-CA source measurements to within 4\%.
In the Far South, the two more recent ATCA campaigns show an offset, with the 
AT20G pipeline flux densities being on average $\sim$20\% higher that the
PMN-CA values (Figure 4, right panel).  The CRATES-CA flux densities measured
in December 2005 are also $\sim$10\% higher on average than PMN-CA. The reason
for these differences is not completely clear but may be related to the different
flux-measurement techniques used by the various groups.  In particular,
the AT20G pipeline uses a novel triple correlation (phase closure) method
to measure flux densities rather than making image-based measurements.

\ We considered the possibility that the VLA in the ``A" configuration resolves out 
diffuse flux on $\sim$$7-25\arcsec$ scales accessible to the ATCA ``6A" 
(CRATES-CA) and ``1.5~km" (AT20G) configurations. One check was to examine a
representative sample of CLASS sources imaged with FIRST \citep{first}, a large 
northern survey at 1.4~GHz with the VLA in the ``B" configuration, sensitive on 
$\sim$$4-120\arcsec$ scales. These images suggest that if the structure
were frequency-independent, 8.4~GHz ``A"-array VLA observations would resolve out
up to $\sim$15\% of the flux density for $\sim$10\% of sources and $\sim$5\% 
of the flux density for another $\sim$10\% of the sources. The over-resolved
sources were, as expected, dominated by objects with core plus double-lobe
morphology at 1.4~GHz. Of course, this extended structure generally has
quite a steep spectrum and should contribute substantially less to the full 
source flux density at 8.4~GHz.  We conclude that over-resolution may 
contribute up to 2\% of the average flux density-scale discrepancy and may 
indeed be more important for the AT20G data.  However, it cannot fully explain
the discrepancy, especially since the PMN-CA and VLA overlap sources show good
agreement in the flux densities.
In any case, we apply the flux density scalings listed in Table 4 to bring
all the surveys to a common scale before defining the final CRATES 8.4~GHz
sample.

\begin{figure}[ht]
\centering
\plottwo{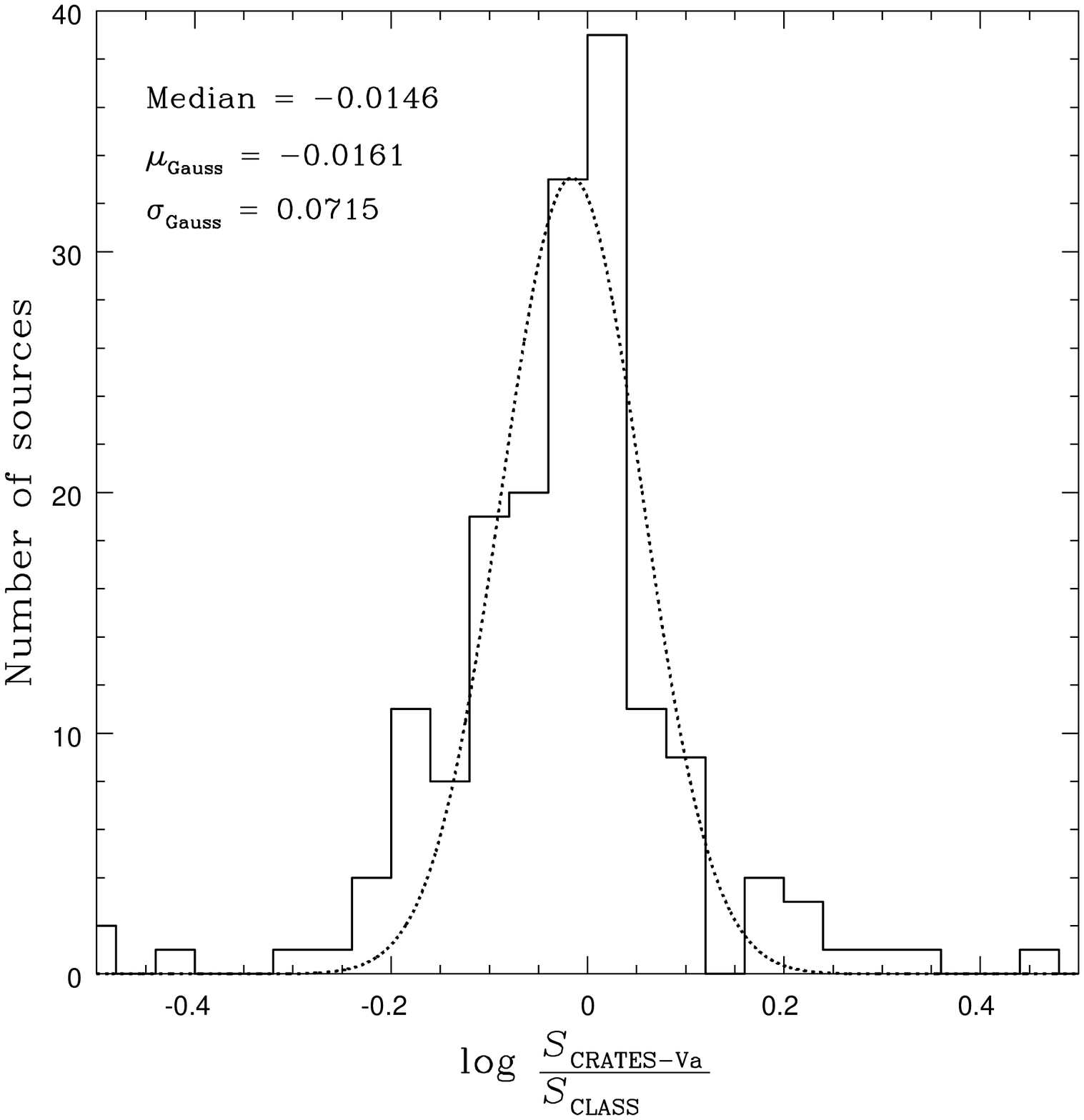}{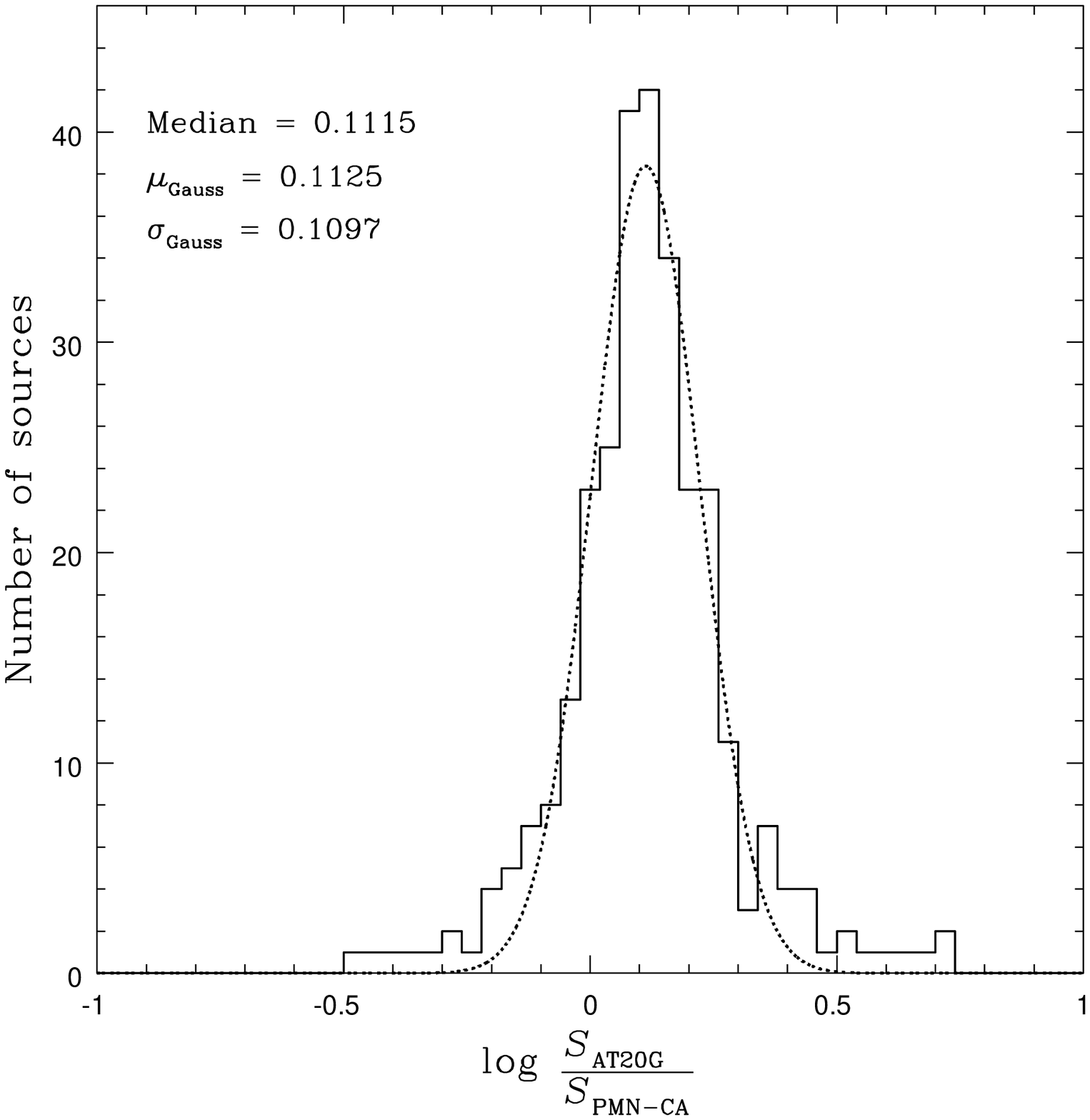}
\caption[highfreq]{Radiometric cross-comparisons for the VLA and for ATCA. \\
{\it Left:} Comparison of CRATES-Va flux density to CLASS flux density for 170 sources. \\
{\it Right:} Comparison of AT20G flux density to PMN-CA flux density for 294 sources.}
\label{highfreq_01}
\end{figure}

\subsection{Final CRATES Catalog}

\ As noted above, many sources were observed at multiple epochs, either
within the same survey or across surveys. We identify components 
between epochs, starting with the brightest component,
by requiring $0\,\farcs25$ matches for the VLA ``A"-array data and 
$1\,\farcs5$ matches for the CA 6~km-array observations. After excluding,
when possible, the AP0282 measurements (see above), we estimate final average 
component flux densities by averaging across all epochs. The fiducial position 
of each component is taken to be that at its brightest epoch. Each source thus has an 
``epoch-averaged" set of components. We have examined all complex sources
and pruned the catalog of obvious errors, such as negative components stemming
from mapping errors. There may also be small negative components fit to PSF residuals
around bright sources. In these cases, we have deleted the negative components
and also pruned the map of all components with a smaller absolute flux density value
that are closer
to the principal component. This serves to suppress insignificant complex structure
in images that are dominated by bright sources and are therefore dynamic range-limited.

\ The first page of the resulting catalog
is shown in Table 5. We also provide associations between
the X-band detections and sources from the low-frequency catalogs, NVSS and
SUMSS; this allows us to report a new, fully interferometric (but non-simultaneous) 
spectral index for the low-frequency sources. Note that while we have found the
low-frequency position valuable in flagging map errors in multi-epoch observations,
we have made no attempt to flag single-epoch sources with discrepant positions.
Such discrepancies may naturally arise when large-scale steep-spectrum emission
dominates the low frequency flux. We do, however, tabulate the associated low-frequency
position, and users may find that for bright, flat-spectrum single-component sources,
such positional disagreement may be a useful pointer to questionable 8~GHz astrometry.
The full table is available in electronic form.

\ From the areal densities in Table 1, we can estimate how close we have come to
a uniform $|b|>10^\circ$ survey. With the shallower S5 4.8~GHz survey in the Far
North, we are missing about 2/3 ($\sim$160) of the expected sources. The PMN holes
in the equatorial zone cover about 300 square degrees, and so another $\sim$100 sources
are missing from these regions. Finally, the 30 square degrees around the southern pole, 
also missing from PMN, should contain another $\sim$10 sources. Thus, overall, we have
covered $\sim$11100/(11100+270) = 97.6\% of the anticipated bright, high-latitude,
flat-spectrum sources.

\begin{deluxetable}{ccccccc}
\tabletypesize{\small}
\setlength{\tabcolsep}{0.15in}
\tablewidth{0pt}
\tablecaption{Pairwise comparisons of X-band survey flux densities.}
\tablehead{
  \colhead{Link}&
  \colhead{First survey}&
  \colhead{Second survey}&
  \colhead{$N$\tablenotemark{a}}&
  \colhead{$\mu$\tablenotemark{b}}&
  \colhead{$\sigma_\mu$\tablenotemark{c}}&
  \colhead{$\mu / \sigma_\mu$}}

\startdata
1&CRATES-Va&CLASS&170&$-0.0161$&0.00548&$-2.94$\\
2&CRATES-V1&CRATES-Va&56&$-0.0178$&0.01986&$-0.89$\\
3&CRATES-V2&CRATES-Va&310&$+0.0191$&0.00704&$+2.71$\\
4&CRATES-V3&CRATES-Va&16&$-0.0073$&0.00740&$-0.99$\\
5&PMN-CA&CRATES-Va&39&$-0.0081$&0.01294&$-0.63$\\
6&AT20G&CRATES-Va&59&$+0.0976$&0.01521&$+6.42$\\
7&AT20G&PMN-CA&294&$+0.1125$&0.00640&$+17.6$\\
8&CRATES-CA&AT20G&52&$-0.0387$&0.01048&$-3.69$\\
\enddata

\tablenotetext{a}{Number of sources common to the first survey and
the second survey.}
\tablenotetext{b}{Mean of the Gaussian fit to the distribution of
$\log (S_{\mathrm {first}}/S_{\mathrm {second}})$.}
\tablenotetext{c}{Standard error in the determination of $\mu$.}

\end{deluxetable}

\begin{deluxetable}{crcc}
\tabletypesize{\small}
\setlength{\tabcolsep}{0.15in}
\tablewidth{0pt}
\tablecaption{Flux density correction factors for each X-band survey.}
\tablehead{
  \colhead{Survey}&
  \colhead{Chain\tablenotemark{a}}&
  \colhead{CF\tablenotemark{b}}&
  \colhead{$\sigma_\mathrm{CF}$\tablenotemark{b}}}

\startdata
CRATES-Va&1 $\rightarrow$ CLASS&1.038&0.013\\
CRATES-V1&2 $\rightarrow$ 1 $\rightarrow$ CLASS&1.038&0.013\\
CRATES-V2&3 $\rightarrow$ 1 $\rightarrow$ CLASS&0.993&0.020\\
CRATES-V3&4 $\rightarrow$ 1 $\rightarrow$ CLASS&1.038&0.013\\
PMN-CA&5 $\rightarrow$ 1 $\rightarrow$ CLASS&1.038&0.013\\
AT20G&7 $\rightarrow$ 5 $\rightarrow$ 1 $\rightarrow$ CLASS&0.816&0.024\\
CRATES-CA&8 $\rightarrow$ 7 $\rightarrow$ 5 $\rightarrow$ 1 $\rightarrow$ CLASS&0.892&0.035\\
\enddata

\tablenotetext{a}{See Table 3 for the details of each link in the chain.}
\tablenotetext{b}{\phantom{a} $\!\!\!\!\!\!\!$ Links with $|\mu| < 2\sigma_\mu$ are bypassed; see text.}

\end{deluxetable}

\input{stub.tab5.tex}

\input{stub.tab6.tex}

\section{Additional Observations}

\ As noted above, there are gaps in the 4.8~GHz surveys, which are the parents of CRATES.
We attempted to select bright, flat-spectrum sources from lower-frequency catalogs
to identify those most like the true CRATES sources.  In the Far North, we selected 90 such
sources by comparing the NVSS and 0.33~GHz WENSS survey \citep{wenss}, and 17 were observed.
For the PMN holes in the Equatorial South, we used NVSS and the 0.385~GHz Texas survey
\citep{texas} to select 107 sources, of which 56 were observed.  Many were indeed bright,
compact sources at 8.4\,GHz
and thus are likely blazar counterparts. The first page of results, treated with the
same NVSS matching method as CRATES proper, is shown in Table 6.  Post facto
4.8\,GHz single-dish flux densities could, in principle, bless these as fully
equivalent to the CRATES sources. Nominally, these additional sources should
bring the high-latitude completeness to $>$98\%.  The full table of results is available
in electronic form.

\section{Value-Added Data Products}

\subsection{WMAP Point Sources}

\ The WMAP three-year data release included a catalog of 323 point sources \citep{wmap}
detected in the sky maps in some or all of the WMAP frequency bands (K~=~23~GHz, Ka~=~33~GHz,
Q~=~41~GHz, V~=~61~GHz, and W~=~94~GHz).
This shallow but uniform catalog offers an opportunity for us to check how well
CRATES does at selecting bright flat-spectrum core sources. We expect the majority
of the WMAP point sources to be in our survey. Indeed, only 38 WMAP point sources do not have
counterparts in CRATES.  Of these, 6 are in the Galactic plane and thus excluded
{\it a priori}.  Four other sources have faint ($<$100~mJy) 4.8~GHz associations and are
low-significance WMAP sources, so the reality of the WMAP detections is questionable.
The remainder of the missing objects are dominated by sources with very high flux 
densities ($S_{4.85 \mathrm{\:GHz}} \ga$ 1~Jy) but with spectra too steep for
admission into CRATES (i.e., $\alpha_{4.85/1.4}< -0.5$).

\ Figure 5 shows a scatter plot of the spectral index between 4.85~GHz and WMAP
Q-band (41~GHz) vs.\ the spectral index between 4.85~GHz and NVSS or SUMSS. Sources
to the left of the vertical dashed line are excluded from CRATES. Sources in the 
lower left corner, near the diagonal line, have steep spectra both at low and high
frequencies. These are appropriately excluded from CRATES and are indeed present
in the WMAP catalog only because they are very bright (generally low-$z$) sources.
There are, however, a handful of sources missing from CRATES near 
$\alpha_{4.85/41} \approx 0$. In these cases, true flat-spectrum cores are apparently
swamped at low frequencies by extended steep-spectrum emission; inspection of FIRST images
of these sources, where available, supports this conclusion.  These missing
sources are again bright.  Specifically, 19\% of the WMAP sources with 
$S_{4.85 \mathrm{\:GHz}} > 2$~Jy fail our CRATES cut while 3\% of fainter sources 
(median flux density $\sim$1~Jy) fail. If this trend holds,
and the fraction of WMAP sources missed by our CRATES cut continues to decrease at
lower flux densities, then
we conclude that less than 1\% of the bright ($>$65~mJy) flat-spectrum
compact cores are missing from the CRATES sample, which is dominated by $\sim$0.1~Jy sources.

\begin{figure}[ht]
\begin{center}
\includegraphics[width=0.45\textwidth]{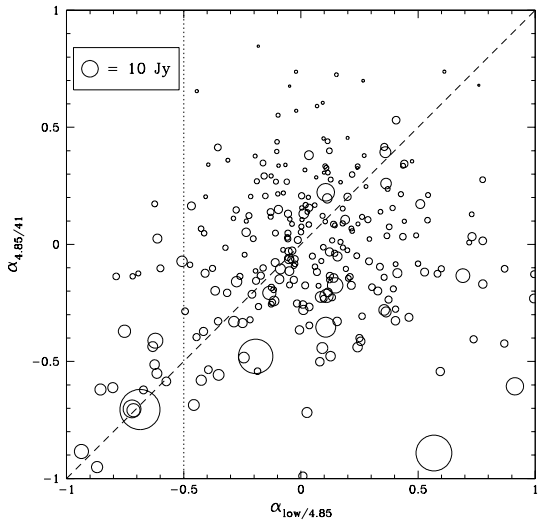}
\caption{Comparison of high-frequency spectral indices to low-frequency spectral indices for
WMAP sources.  $\alpha_\mathrm{low/4.8}$ is computed between NVSS or SUMSS, depending on
declination, and 4.85~GHz.  The size of the circle is proportional to the square root of the
flux density at 4.8~GHz.  The circle for a 10~Jy source is shown for comparison.  Sources
in the lower right are gigahertz-peaked sources.}

\label{highfreq_02}
\end{center}
\end{figure}

\subsection{Component Analysis}

\ To complete the description of the CRATES catalog, we have made a crude 
morphological classification of the sources using the maps and model fit components.
In order of decreasing compactness, we have classified the sources as
``P" = point source, ``S" = short jet ($\le$$1\arcsec$ for VLA maps), ``L" = long jet
($>$$1\arcsec$ separation), ``D" = double (component flux density ratio $\le$$2\times$),
and ``C" = complex (see Figure 6).  Using all components brighter than 1~mJy and separations 
$0\,\farcs 125 - 6\arcsec$, two-component sources were classified according to the criteria 
above.  Long jet (``L") sources were further required to have a core/jet 
flux density ratio $\le$100 to avoid tagging side-lobes as jets. We also found that
very compact VLA doubles ``D" with separation less than $3\times$ our minimum component
separation (i.e., $0\,\farcs375$) were often produced by poor PSF fitting;
such sources are re-classified as core-dominated ``P."
Note that all epoch-averaged components are retained in the catalog, even if they
were flagged for exclusion in the classification exercise.
We inspected all sources with three or more components and placed them
into one of the categories. Visual inspection of a sample of the two-component
sources shows that the numerical cuts above select the appropriate category with
$\sim$5\% misclassification. Only 61 sources showed multiple components in
the ATCA data; all were classified by hand. These classifications are shown in the first line
for each target (following the interferometric spectral index) in the CRATES catalog, Table 5.
Sources classified as ``N" were observed at X-band but not detected.
We show the breakdown of source classifications and mean spectral indices 
for the relatively uniform VLA observations ($\delta > -40^\circ$) in
Table 7.

\begin{figure}[ht]
\begin{center}
\includegraphics[width=0.99\textwidth]{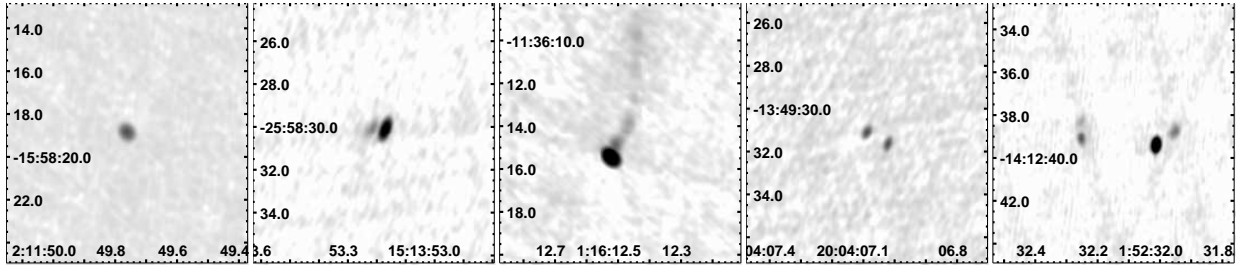}
\caption{Sample images of our five source classes. The images are drawn from our new
VLA observations and show, from left to right, a log intensity grayscale map of
sources with decreasing compactness: ``P", ``S", ``L", ``D", and ``C."
}
\label{Samplemaps}
\end{center}
\end{figure}
\begin{deluxetable}{llrc}
\tabletypesize{\small}
\setlength{\tabcolsep}{0.15in}
\tablewidth{0pt}
\tablecaption{Morphological classifications for VLA observations.}
\tablehead{
  \multicolumn{2}{c}{Morphological class}&
  \colhead{Fraction}&
  \colhead{$\langle \alpha_{\mathrm{low}/8.4} \rangle$}}

\startdata
``N"&No detection&6.0\%&N/A\\
``P"&Point source&84.3\%&$-0.177\pm0.008$\\
``S"&Short jet&5.1\%&$-0.444\pm0.028$\\
``L"&Long jet&3.6\%&$-0.428\pm0.035$\\
``D"&Double&0.6\%&$-0.691\pm0.068$\\
``C"&Complex&0.4\%&$-0.444\pm0.155$\\
\enddata
\end{deluxetable}

\section{Conclusions}

	We have assembled a large and nearly uniform catalog of interferometric
observations of bright flat-spectrum radio sources. This list should be particularly
valuable for statistical comparison with other all-sky surveys. We should remember 
that the GB6 parent catalog of this survey is now some 20~y old. Given that these 
sources are significantly variable, we would not be surprised if some new bright 
flat-spectrum sources have appeared in the interim. However, this survey is 
the most complete list of such objects at present.  Since such
flat-spectrum sources are argued to dominate the source population at both
microwave and GeV energies \citep{giommi}, this catalog will be especially
useful for comparison with the forthcoming Planck and GLAST/LAT sky surveys.
Planck should have a sensitivity $\ge$30$\times$ that of WMAP for point
sources, with a typical positional uncertainty of $5-10\arcmin$, and should
detect several thousand sources at $5\sigma$ significance.
Similarly, the LAT sky survey of the GLAST mission is expected to
detect 1,000$-$10,000 blazars \citep{cm98}. \citet{srm} have shown that
core flux density and spectral index can be used to select likely counterparts of
$\gamma$-ray blazars, so this survey should be ideal for application to
the GLAST/LAT survey detections.  Indeed, we have selected a $\sim$15\% subset of
this survey with the greatest similarity (in terms of flux density, spectral index, and
X-ray emission) to the {\it EGRET} blazars
and are completing optical identifications of this sample (CGRaBS, the {\bf C}andidate
{\bf G}amma-{\bf Ra}y {\bf B}lazar {\bf S}urvey; Healey et al.\ 2007). 

	These interferometric observations provide the sub-arcsecond positions
needed to make reliable optical associations to faint ($R \sim 23$) magnitudes.
For the VLA observations, we also have good sub-arcsecond scale structure for many
sources, which allows some basic morphological classification. As expected for
such a flat-spectrum sample, most sources are unresolved point sources (``P"). 
These compact sources provide a convenient, relatively bright set of 
potential atmospheric phase fluctuation calibrators for, e.g., ALMA, CARMA, 
MERLIN, and the VLBA. In the $-40^\circ < \delta < 75^\circ$ region, our ``P"
sources with $S_{8.4\:\mathrm{GHz}} \ge 50$\,mJy provide $0.217/\sq^\circ$ potential
calibrators.  We also have a significant population of 
short (``S," $\le$$1\arcsec$) jet emission sources, while the number of
more complex sources is small. The ATCA data also show a handful of ``D"-, ``L"-,
and ``C"-class sources, but most are de facto ``P." The areal density of
potential compact sources in this region is $0.225/{\rm deg^2}$,
but of course these are only known to be compact at the $\sim$$2\arcsec$ scale.
In general, we expect the most compact sources to have the highest
(flattest) spectral indices and the largest variability. A crude measure
of the component compactness or core domination can be inferred from
$S_{\mathrm{max}}/\Sigma S$, where $S_{\mathrm{max}}$ is the flux density of the brightest
component and $\Sigma S$ is the sum of the flux densities of all components.  In the left
panel of Figure 7, we see that $\alpha$ does indeed increase as one moves from ``D" to
``P" sources. However, the few complex ``C" sources have a large range of $\alpha$ values
and a larger uncertainty in the mean.

In the right panel, the trend is also clear, but we see that some complex,
double, and extended jet sources lie above the main distribution, to the
upper left. Interestingly, about 1/2 of known lenses \citep{cllens,wlens02} in CRATES lie in
this region of extended structure but overall flat spectrum, so 
we infer that in some cases flat-spectrum ``C" sources are created by multiple
imaging of compact cores.

\begin{figure}[ht]
\centering
\plottwo{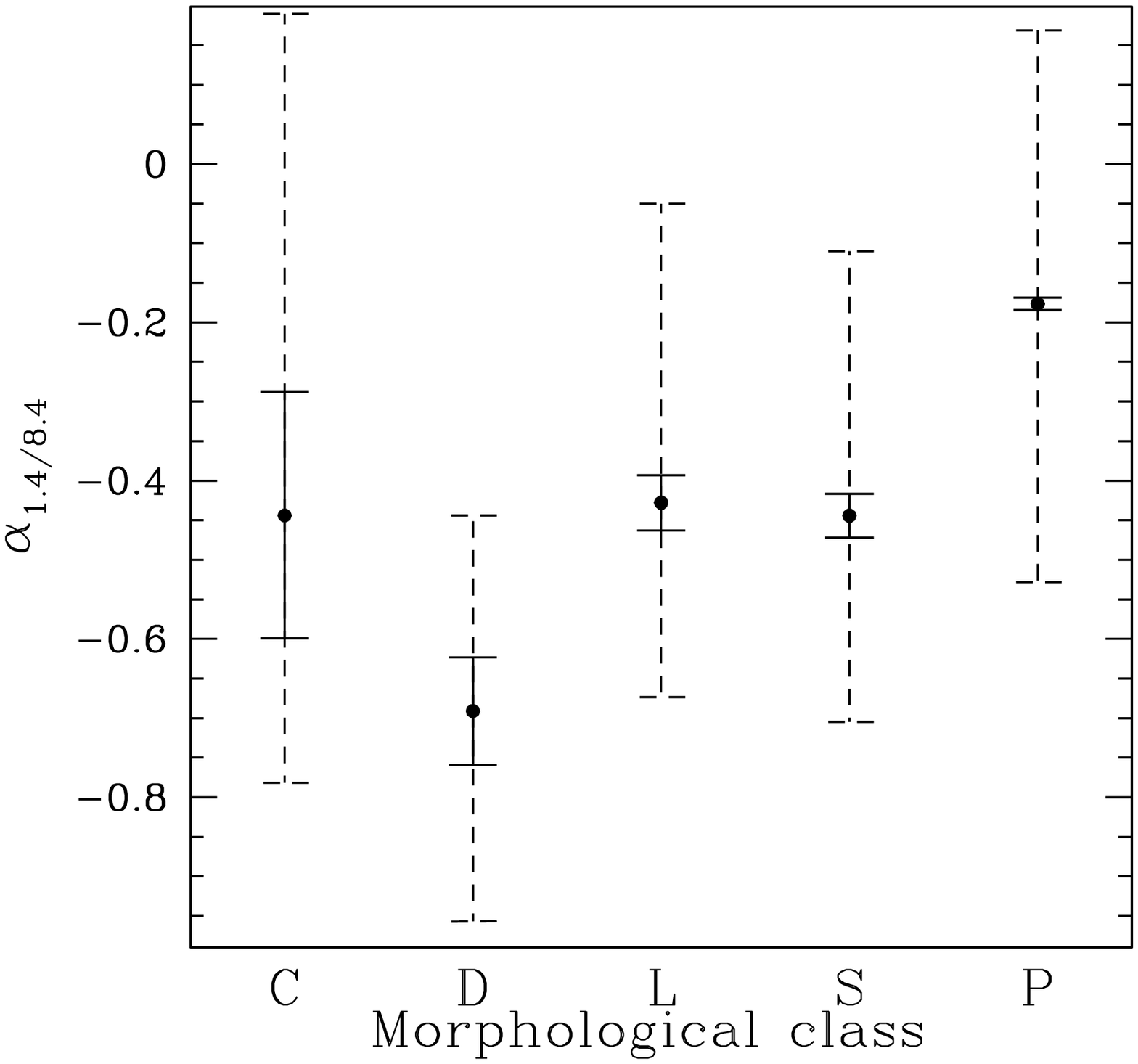}{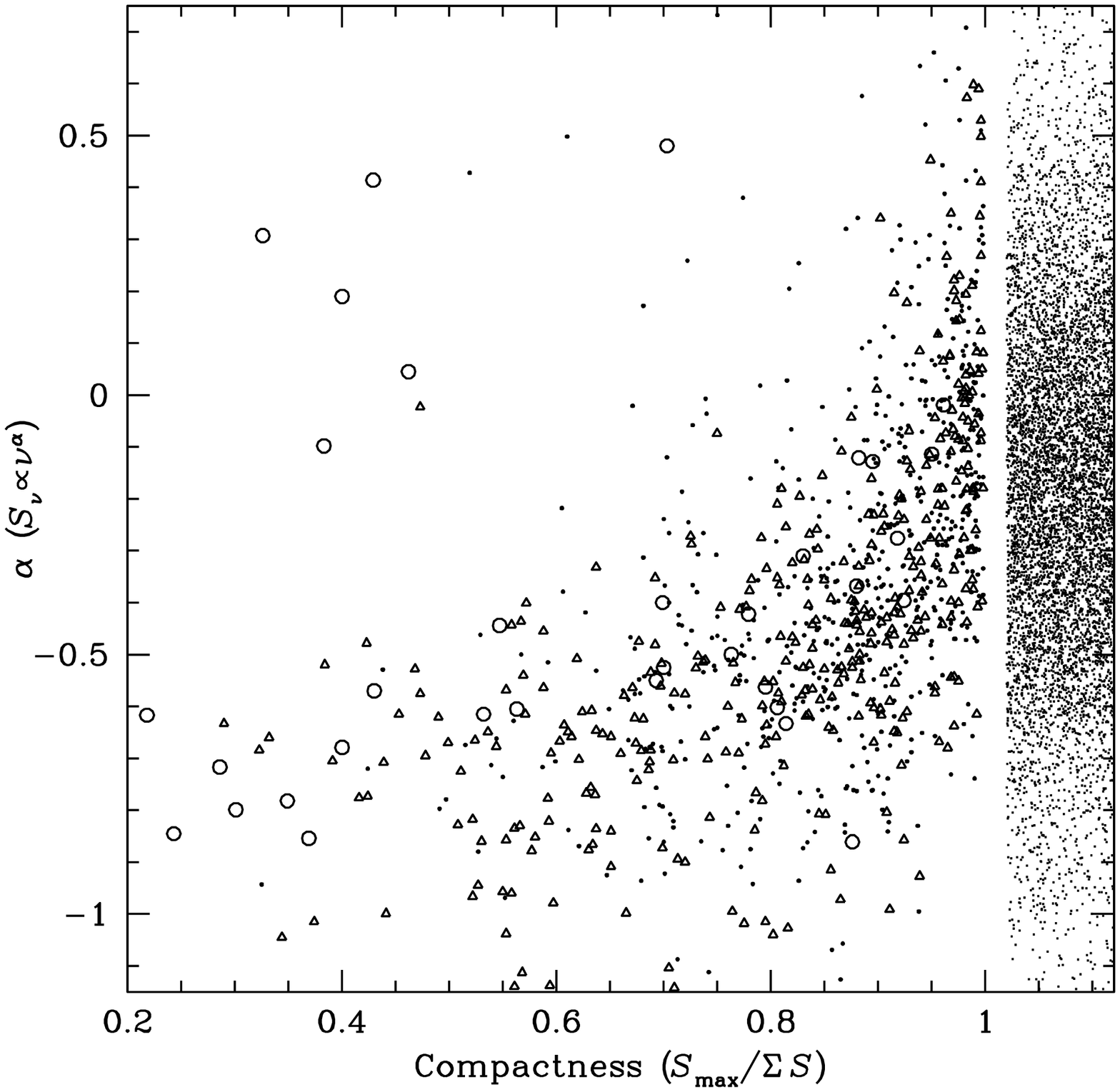}

\caption[compalpha]{Spectral index $\alpha_\mathrm{low/8.4}$ as a function of 
morphological class.\\
{\it Left:} the median $\alpha$ value for each class plotted with the error bar 
for its uncertainty and the 68\% range of spectral indices in the class (dashed 
error bars). There is a general trend toward
flatter spectra as the sources are increasingly core-dominated. \\

{\it Right:} spectral index as a function of compactness (core domination). Point 
sources (``P") and
short jets (``S") are shown as dots, long jets (``L") and extended doubles (``D") as triangles,
and complex (``C") sources as circles. Unresolved (``P") sources are shown in the band at right. 
The trend toward increasing $\alpha$ at increased compactness is clear. The handful
of non-compact, flat-spectrum sources has a large fraction of known gravitational
lenses.}
\label{compalpha}
\end{figure}

	We have also probed the variability of our sample, characterizing
the source variability as that of its brightest component: 
$\mathrm{RMS}(S_i)/\langle S_i \rangle$. Our sampling of this variability 
is highly non-uniform, with epoch separations spanning $\sim 1-10$ y.
The mean variability of unresolved 
cores with multi-epoch observations was 14\%. While the mean variability
of the short jets was 12\%, the number of multi-epoch measurements was
too small to discern clear differences. Interestingly, the few doubles and
complex sources with multi-epoch data showed rather high variability of
20-30\%. This may, however, be a selection effect as a large number of
gravitational lenses and lens candidates are in these classes; these
may have attracted more follow-up observations in the extended lens surveys
and are more likely to be measurably variable.

	The primary purpose of the CLASS survey was to discover gravitational 
lenses, and 22 were found in the $\sim$12,000 sources observed \citep{cllens}. 
We have included $\sim$4800 sources from the CLASS set and have independently 
recognized 
several gravitational lenses in our map inspections, especially among the ``C"
sources. Similarly, gravitational lens searches were the main focus of most
of the CRATES-Va projects, and at least four gravitational lenses have been
identified from a sample of 4000 sources observed in these programs \citep{wlens02}.
Fifteen published lenses from these two survey sets are in the CRATES sample.
As it happens, only $\sim$300 sources north of $\delta = -40^\circ$ were uniquely
observed in our new VLA observations CRATES-V1, -V2, -V3. Thus, scaling from
the CLASS discovery rate ($\sim$1 lens / 550 flat-spectrum sources), we would
not expect any new lens discoveries in this data set. However, the
CRATES-Va reprocessing and the multi-epoch comparison of some sources have
produced improved maps and component measurements. During our 
spectral analysis and classification exercise, we have flagged a number 
of lens candidates for which we will
be pursuing follow-up observations. We have collected initial maps of a
larger fraction of the Far Southern sources, but since the resolution 
($\sim 1\arcsec$ for the maximum 6~km baseline) is not really high enough for
lens searches, it is not surprising that only a few show sufficient resolved 
structure to be worthy of further study.

     In summary, we have assembled a large, targeted cm-wavelength survey 
of compact flat-spectrum sources, covering the full extra-Galactic sky. 
While many of the observations used in this catalog were mined from the
VLA and ATCA archives, over 40\% of the positions and flux densities 
are published here for the first time.  We have attempted to make the 
sensitivity, flux density scale, and astrometric quality as uniform as 
possible.  We expect that upcoming all-sky surveys, especially
the GLAST/LAT and Planck surveys, will find this catalog an excellent
list for making statistical association with flat-spectrum counterparts and
for large-sample studies of their multiwavelength properties. We
expect that blazar studies will particularly benefit from this large,
uniform survey. We are now focusing on the best candidates for high-energy
blazar emitters and collecting multiwavelength associations and spectral
IDs. While there are other important blazar classification efforts underway
(e.g., Massaro et al.\ 2005), we hope that the extent and uniformity of the CRATES
sample will make statistical studies with these objects particularly powerful.

\acknowledgements
     The National Radio Astronomy Observatory is operated by Associated Universities, Inc.,
under cooperative agreement with the National Science Foundation.  SEH was supported by SLAC
under DOE contract DE-AC03-76SF00515.  We thank D.~Sowards-Emmerd, F.~Heatherington and
C.~M.~L.~Williams for assistance with the early phases of the data reduction.

\end{document}

%% file: stub.tab5.tex
\begin{deluxetable}{rrrcrrcrcrrrrc}
\setlength{\tabcolsep}{0.03in}
\tabletypesize{\scriptsize}
\tablewidth{0pt}
\tablecaption{The CRATES catalog.}
\tablehead{
 \colhead{4.8 GHz name\tablenotemark{a,{\rm b}}}&
 \colhead{$S_{4.8}$\tablenotemark{b}}&
 \colhead{$\alpha_{\mathrm{low}/4.8}$}&
 &
 \multicolumn{2}{c}{8.4 GHz position\tablenotemark{a}}&
 \colhead{Or.\tablenotemark{c}}&
 \colhead{$S_{8.4}$}&
 &
 \multicolumn{2}{c}{Low freq. position\tablenotemark{a,{\rm d}}}&
 \colhead{$S_{\mathrm{low}}$\tablenotemark{d}}&
 \colhead{$\alpha_{\mathrm{low}/8.4}$}&
 \colhead{Morph.}\\

 &
 \colhead{(mJy)}&
 &
 &
 \colhead{RA}&
 \colhead{DEC}&
 &
 \colhead{(mJy)}&
 &
 \colhead{RA}&
 \colhead{DEC}&
 \colhead{(mJy)}&
 &
 \colhead{class\tablenotemark{e}}}

\startdata
J000000$-$002157&116&$-$0.498&\phantom{Q}&00 00 01.66&$-$00 22 10.0&V&89.5&\phantom{Q}&00 00 01.66&$-$00 22 09.8&215.4&$-$0.490&P\\
J000004$-$135133&74&0.142&\phantom{Q}&00 00 03.13&$-$13 52 00.9&V&36.8&\phantom{Q}&00 00 03.09&$-$13 52 00.0&62.0&$-$0.291&P\\
J000018$+$024812&65&$-$0.172&\phantom{Q}&00 00 19.28&$+$02 48 14.7&V&85.2&\phantom{Q}&00 00 19.27&$+$02 48 14.7&80.5&0.032&P\\
J000019$-$853946&98&0.073&\phantom{Q}&00 00 12.04&$-$85 39 19.9&A&66.3&\phantom{Q}&00 00 11.43&$-$85 39 20.1&102.9&$-$0.192&P\\
J000021$-$322118&535&$-$0.005&\phantom{Q}&00 00 20.40&$-$32 21 01.2&V&279.8&\phantom{Q}&00 00 20.33&$-$32 20 59.1&538.1&$-$0.365&P\\
J000026$+$030706&91&0.249&\phantom{Q}&00 00 27.02&$+$03 07 15.6&V&100.7&\phantom{Q}&00 00 27.03&$+$03 07 16.4&66.8&0.229&P\\
J000035$+$291424&96&0.165&\phantom{Q}&00 00 35.13&$+$29 14 35.8&V&64.7&\phantom{Q}&00 00 35.09&$+$29 14 35.0&78.2&$-$0.106&P\\
J000040$+$391758&140&$-$0.332&\phantom{Q}&00 00 41.53&$+$39 18 04.2&V&97.3&\phantom{Q}&00 00 41.51&$+$39 18 04.4&211.6&$-$0.323&L\\
&&&\phantom{Q}&00 00 41.49&$+$39 18 05.1&V&21.4&\phantom{Q}&&&&&\\
J000044$+$030744&92&0.475&\phantom{Q}&00 00 44.33&$+$03 07 54.2&V&63.3&\phantom{Q}&00 00 44.31&$+$03 07 54.3&51.0&0.121&P\\
J000046$-$392352&75&$-$0.497&\phantom{Q}&00 00 46.06&$-$39 22 34.2&A&23.2&\phantom{Q}&00 00 46.13&$-$39 22 37.1&139.1&$-$1.000&P\\
J000048$+$121810&78&$-$0.324&\phantom{Q}&&&&&\phantom{Q}&&&&&N\\
J000103$-$294013&93&$-$0.302&\phantom{Q}&00 01 07.73&$-$29 40 32.9&V&55.9&\phantom{Q}&00 01 07.72&$-$29 40 29.8&135.3&$-$0.493&P\\
J000104$-$370321&79&2.268&\phantom{Q}&&&&&\phantom{Q}&&&&&N\\
J000105$-$155101&305&$-$0.121&\phantom{Q}&00 01 05.33&$-$15 51 07.1&V&335.9&\phantom{Q}&00 01 05.26&$-$15 51 07.0&354.3&$-$0.030&P\\
J000109$+$191428&233&$-$0.119&\phantom{Q}&00 01 08.62&+19 14 33.8&V&504.2&\phantom{Q}&00 01 08.63&+19 14 34.2&270.0&0.349&P\\
J000113$-$345124&81&0.479&\phantom{Q}&00 01 12.45&$-$34 51 52.1&V&34.7&\phantom{Q}&00 01 12.41&$-$34 51 54.6&44.7&$-$0.141&P\\
J000114$+$235801&121&$-$0.215&\phantom{Q}&00 01 14.86&+23 58 10.6&V&132.7&\phantom{Q}&00 01 14.85&+23 58 10.4&158.0&$-$0.097&P\\
J000117$-$074633&148&$-$0.298&\phantom{Q}&00 01 18.03&$-$07 46 27.0&V&116.2&\phantom{Q}&00 01 18.00&$-$07 46 26.8&214.2&$-$0.341&P\\
J000119$+$474202&135&$-$0.478&\phantom{Q}&00 01 19.04&$+$47 42 00.7&V&100.5&\phantom{Q}&00 01 19.06&$+$47 42 00.7&244.5&$-$0.413&P\\
&&&\phantom{Q}&00 01 19.32&$+$47 42 05.3&V&5.3&\phantom{Q}&&&&&\\
&&&\phantom{Q}&00 01 19.59&$+$47 42 00.5&V&3.8&\phantom{Q}&&&&&\\
&&&\phantom{Q}&00 01 18.14&$+$47 41 45.8&V&3.1&\phantom{Q}&&&&&\\
&&&\phantom{Q}&00 01 18.33&$+$47 42 01.1&V&2.7&\phantom{Q}&&&&&\\
&&&\phantom{Q}&00 01 17.69&$+$47 41 55.6&V&1.3&\phantom{Q}&&&&&\\
J000121$+$444025&165&$-$0.455&\phantom{Q}&00 01 21.38&$+$44 40 27.2&V&67.6&\phantom{Q}&00 01 21.39&$+$44 40 27.3&290.5&$-$0.515&S\\
&&&\phantom{Q}&00 01 21.36&$+$44 40 27.2&V&26.4&\phantom{Q}&&&&&\\
&&&\phantom{Q}&00 01 21.36&$+$44 40 27.0&V&19.4&\phantom{Q}&&&&&\\
&&&\phantom{Q}&00 01 22.67&$+$44 40 21.2&V&1.4&\phantom{Q}&&&&&\\
&&&\phantom{Q}&00 01 21.31&$+$44 40 28.1&V&0.7&\phantom{Q}&&&&&\\
J000122$+$252655&73&0.078&\phantom{Q}&00 01 21.67&$+$25 26 55.5&V&41.3&\phantom{Q}&00 01 21.66&$+$25 26 55.6&66.3&$-$0.264&P\\
J000122$-$250010&102&0.287&\phantom{Q}&00 01 22.67&$-$25 00 18.8&V&40.8&\phantom{Q}&00 01 22.67&$-$25 00 18.7&71.4&$-$0.312&P\\
J000124$-$065618&116&0.525&\phantom{Q}&00 01 25.59&$-$06 56 25.0&V&77.2&\phantom{Q}&00 01 25.53&$-$06 56 24.9&60.4&0.137&P\\
J000129$+$435205&94&0.066&\phantom{Q}&00 01 29.13&$+$43 51 56.1&V&64.1&\phantom{Q}&00 01 29.09&$+$43 51 55.9&86.6&$-$0.146&P\\
&&&\phantom{Q}&00 01 28.55&$+$43 52 01.8&V&1.5&\phantom{Q}&&&&&\\
&&&\phantom{Q}&00 01 28.72&$+$43 52 04.8&V&1.1&\phantom{Q}&&&&&\\

\enddata

\tablenotetext{a}{J2000 position.}
\tablenotetext{b}{From S5, GB6, or PMN.}
\tablenotetext{c}{Origin of 8.4 GHz position: V = VLA, A = ATCA, X = VLA Program AP0282; see text.}
\tablenotetext{d}{From NVSS or SUMSS.}
\tablenotetext{e}{Morphological classification: ``N" = no detection, ``P" = point source, ``S" = short jet, ``L" = long jet, \\
\hspace{120.5pt}``D" = double, ``C" = complex morphology.  See text for full description.}

\end{deluxetable}

%% file: stub.tab6.tex
\begin{deluxetable}{rrrcrrcrcrrrrc}
\setlength{\tabcolsep}{0.03in}
\tabletypesize{\scriptsize}
\tablewidth{0pt}
\tablecaption{Results of additional observations.}
\tablehead{
 \colhead{Pointing name\tablenotemark{a,{\rm b}}}&
 \colhead{$S_{1.4}$\tablenotemark{b}}&
 \colhead{$\alpha_{\mathrm{low}/1.4}$\tablenotemark{c}}&
 &
 \multicolumn{2}{c}{8.4 GHz position\tablenotemark{a}}&
 \colhead{Or.\tablenotemark{d}}&
 \colhead{$S_{8.4}$}&
 &
 \multicolumn{2}{c}{NVSS position\tablenotemark{a}}&
 \colhead{$S_{\mathrm{NVSS}}$}&
 \colhead{$\alpha_{1.4/8.4}$}&
 \colhead{Morph.}\\

 &
 \colhead{(mJy)}&
 &
 &
 \colhead{RA}&
 \colhead{DEC}&
 &
 \colhead{(mJy)}&
 &
 \colhead{RA}&
 \colhead{DEC}&
 \colhead{(mJy)}&
 &
 \colhead{class\tablenotemark{e}}}

\startdata

J012808$+$792846&150&0.160&\phantom{Q}&01 28 08.88&$+$79 28 46.1&W&79.3&\phantom{Q}&01 28 08.93&$+$79 28 46.1&149.5&$-$0.354&P\\
J013829$+$861140&101&0.759&\phantom{Q}&01 38 29.66&$+$86 11 40.9&W&39.7&\phantom{Q}&01 38 29.65&$+$86 11 40.1&101.0&$-$0.521&P\\
J061915$+$800706&155&1.356&\phantom{Q}&06 19 15.34&$+$80 07 06.0&W&31.1&\phantom{Q}&06 19 15.26&$+$80 07 06.2&154.9&$-$0.896&P\\
J072712$+$854517&268&0.021&\phantom{Q}&&&&&&&&&&N\\
J105421$+$862936&234&0.172&\phantom{Q}&10 54 22.27&$+$86 29 36.1&W&177.7&\phantom{Q}&10 54 21.71&$+$86 29 36.4&234.3&$-$0.154&P\\
J123306$+$823350&96&0.444&\phantom{Q}&&&&&&&&&&N\\
J132247$+$760649&112&0.097&\phantom{Q}&13 22 47.39&$+$76 06 47.6&W&36.5&\phantom{Q}&13 22 47.12&$+$76 06 49.3&111.9&$-$0.625&P\\
J140420$+$782950&128&0.808&\phantom{Q}&14 04 20.31&$+$78 29 50.6&W&52.2&\phantom{Q}&14 04 20.32&$+$78 29 50.6&128.3&$-$0.502&P\\
J151032$+$800005&116&0.062&\phantom{Q}&15 10 32.75&$+$80 00 05.3&W&183.6&\phantom{Q}&15 10 32.48&$+$80 00 05.2&116.1&0.256&P\\
J155058$+$815423&97&0.164&\phantom{Q}&15 50 58.29&$+$81 54 24.2&W&28.0&\phantom{Q}&15 50 58.25&$+$81 54 23.8&96.8&$-$0.692&P\\
J183205$+$804941&282&0.414&\phantom{Q}&18 32 05.01&$+$80 49 41.8&W&87.2&\phantom{Q}&18 32 05.27&$+$80 49 41.8&281.5&$-$0.654&P\\
J200640$+$760543&132&0.047&\phantom{Q}&20 06 40.91&$+$76 05 45.3&W&48.1&\phantom{Q}&20 06 40.72&$+$76 05 43.7&131.8&$-$0.562&P\\
J211649$+$762819&130&0.171&\phantom{Q}&21 16 49.29&$+$76 28 20.7&W&44.9&\phantom{Q}&21 16 49.40&$+$76 28 19.6&129.9&$-$0.593&P\\
J215834$+$825348&104&0.969&\phantom{Q}&21 58 34.16&$+$82 53 48.1&W&37.3&\phantom{Q}&21 58 34.32&$+$82 53 48.6&104.2&$-$0.573&P\\
J222721$+$773319&137&0.129&\phantom{Q}&22 27 21.62&$+$77 33 19.2&W&66.9&\phantom{Q}&22 27 21.66&$+$77 33 19.7&136.6&$-$0.398&P\\
J232553$+$835636&119&0.154&\phantom{Q}&23 25 53.01&$+$83 56 37.4&W&64.8&\phantom{Q}&23 25 52.98&$+$83 56 36.9&119.2&$-$0.340&P\\
J232628$+$800813&163&0.906&\phantom{Q}&23 26 28.44&$+$80 08 12.6&W&15.3&\phantom{Q}&23 26 28.37&$+$80 08 13.1&163.4&$-$1.322&P\\
\\[-5pt]
\hline
\\[-5pt]
J120741$-$010637&136&$-$0.544&\phantom{Q}&12 07 41.68&$-$01 06 36.7&T&216.8&\phantom{Q}&12 07 41.66&$-$01 06 37.4&135.5&0.262&P\\
J121031$-$013650&659&$-$0.581&\phantom{Q}&12 10 31.75&$-$01 36 56.1&T&21.8&\phantom{Q}&12 10 31.37&$-$01 36 50.1&659.4&$-$1.698&P\\
&&&&12 10 31.76&$-$01 36 56.2&T&9.7&&&&\\
J121314$-$015904&157&$-$0.677&&&&&&&&&&&N\\
J121432$-$041603&232&$-$0.608&\phantom{Q}&12 14 32.43&$-$04 16 01.8&T&22.8&\phantom{Q}&12 14 32.37&$-$04 16 03.4&232.2&$-$0.905&L\\
&&&&12 14 32.45&$-$04 16 01.8&T&12.2&&&&\\
&&&&12 14 32.41&$-$04 16 01.8&T&5.3&&&&\\
&&&&12 14 32.38&$-$04 16 01.8&T&5.6&&&&\\
J121514$-$062804&366&$-$0.230&\phantom{Q}&12 15 14.39&$-$06 28 03.9&T&417.1&\phantom{Q}&12 15 14.42&$-$06 28 03.6&367.0&0.071&P\\
J121755$-$033721&218&$-$0.666&\phantom{Q}&12 17 55.26&$-$03 37 23.2&T&18.3&\phantom{Q}&12 17 55.30&$-$03 37 21.2&217.7&$-$1.059&S\\
&&&&12 17 55.26&$-$03 37 23.3&T&14.3&&&&\\
J121836$-$063116&419&$-$0.771&\phantom{Q}&12 18 36.18&$-$06 31 16.6&T&120.2&\phantom{Q}&12 18 36.18&$-$06 31 15.9&418.7&$-$0.696&P\\
J122716$-$044533&130&$-$0.641&\phantom{Q}&12 27 16.55&$-$04 45 32.5&T&20.9&\phantom{Q}&12 27 16.53&$-$04 45 32.8&129.9&$-$1.020&P\\
J122852$-$063146&403&$-$0.454&\phantom{Q}&12 28 52.18&$-$06 31 46.6&T&72.0&\phantom{Q}&12 28 52.20&$-$06 31 46.3&402.9&$-$0.662&S\\
&&&&12 28 52.19&$-$06 31 46.5&T&33.8&&&&\\
&&&&12 28 52.21&$-$06 31 46.5&T&17.2&&&&\\
J123029$-$051001&225&$-$0.680&&&&&&&&&&&N\\
J123838$-$065614&186&$-$0.595&\phantom{Q}&12 38 38.49&$-$06 56 12.2&T&184.0&\phantom{Q}&12 38 38.32&$-$06 56 13.5&186.4&$-$0.007&P\\
J124412$-$080605&282&$-$0.683&\phantom{Q}&12 44 12.30&$-$08 06 03.9&T&40.8&\phantom{Q}&12 44 12.32&$-$08 06 04.9&282.5&$-$1.080&P\\

\enddata

\tablenotetext{a}{J2000 position.}
\tablenotetext{b}{From NVSS.}
\tablenotetext{c}{Between WENSS (330 MHz) and NVSS or Texas Survey (365 MHz) and NVSS.}
\tablenotetext{d}{Origin of low-frequency data: W = WENSS, T = Texas Survey.}
\tablenotetext{e}{Morphological classification: ``N" = no detection, ``P" = point source, ``S" = short jet, ``L" = long jet, \\
\hspace{120.5pt}``D" = double, ``C" = complex morphology.  See text for full description.}

\end{deluxetable}